\documentclass[%
 reprint,
 amsmath,amssymb,
 aps,
pra,
]{revtex4-2}

\usepackage[english]{babel}
\usepackage{comment}
\usepackage{graphicx}
\usepackage{dcolumn}
\usepackage{bm}

\usepackage{tikz}
\usepackage{pgf}
\usetikzlibrary{arrows, automata}
\usepackage{hyperref}
\usetikzlibrary{fit,positioning}
\hypersetup{
    colorlinks=true,
    linkcolor=blue,
    urlcolor=blue,
    citecolor=blue
}
\usepackage{physics}
\usepackage{quantikz}
\usetikzlibrary{angles,quotes}

\begin{document}


\title{Entanglement in Quantum Systems Based on Directed Graphs}

\author{Lucio De Simone}
\email{l.desimone3@student.unisi.it}
\affiliation{DSFTA, University of Siena, Via Roma 56, 53100 Siena, Italy}
\affiliation{INFN Sezione di Perugia, 06123 Perugia, Italy}

\author{Roberto Franzosi}
\email{roberto.franzosi@unisi.it}
\affiliation{DSFTA, University of Siena, Via Roma 56, 53100 Siena, Italy}
\affiliation{INFN Sezione di Perugia, 06123 Perugia, Italy}


\date{\today}

\begin{abstract}
We investigate the entanglement properties of quantum states associated with directed graphs. Using a measure derived from the Fubini–Study metric, we quantitatively relate multipartite entanglement to the local connectivity of the graph. 
In \emph{Entanglement in Directed Graph States} (2025), arXiv:2505.10716,
it is demonstrated that the vertex degree distribution fully determines this entanglement measure and remains invariant under vertex relabeling, highlighting its topological character. As a consequence, the measure depends only on the total degree of each vertex, making it independent of the distinction between incoming and outgoing edges. We apply our framework to several specific graph structures, including hierarchical networks, neural network–inspired graphs, full binary tree and linear bridged cycle graphs, demonstrating how their combinatorial properties influence entanglement distribution. These results provide a geometric perspective on quantum correlations in complex systems, offering potential applications in the design and analysis of quantum networks.
\end{abstract}

\maketitle


\section{Introduction}
Entanglement, besides being one of the most historically puzzling properties of quantum mechanics, serves as the primary resource for quantum cryptography, computation, and other quantum-based technologies. Over the past decades, the quantum information community has developed numerous approaches to characterize its rich phenomenology and diverse properties  \cite{guhne_entanglement_2009,horodecki_quantum_2009}. It has been found that entanglement in multipartite states is a significantly more complex concept compared to bipartite states.
The methods for quantifying entanglement in multipartite states are diverse \cite{guhne_entanglement_2009,horodecki_quantum_2009}. The present study will adopt the Entanglement Distance (ED) as an entanglement measure, initially introduced in Ref. \cite{cocchiarella_entanglement_2020}. It is a measure for general multipartite pure states. The ED has been extended to multipartite mixed states \cite{vesperini_entanglement_2023} and has been applied to various intriguing systems \cite{vafafard_multipartite_2022,nourmandipour_entanglement_2021, Vesperini_2023, vesperini_correlations_2023}. Its theoretical foundation lies in the Fubini-Study metric associated with the local-unitary invariant projective Hilbert space, referred to in this context as the Entanglement Metric. The profound geometric implications of this metric have been further explored in Ref. \cite{Vesperini_2024}. In particular, the present study shows that the ED is especially effective in measuring the entanglement of a special class of states, known as graph states, and is also able to highlight their topological properties.

 Graph states are a rich class of multi-partite entangled states \cite{briegel_persistent_2001,hein_multi-party_2004,hein_entanglement_2006,raussendorf_quantum_2012,płodzień2024manybodyquantumresourcesgraph}. They can be described by a number of parameters that is limited compared to the dimension of the full Hilbert space of a system. In quantum information processing, graph states provide a powerful formalism, especially useful in measurement-based quantum computation \cite{PhysRevLett.86.5188}, quantum error correction \cite{PhysRevA.105.042418}, and protocols for secret key sharing \cite{PhysRevA.78.042309}. Like graphs, they display a combinatorial nature. 

 This work is dedicated to the study of directed graph states, which are generalizations of graph states, associated with directed graphs.
Even for directed graphs, a combinatorial approach is necessary to characterize their properties relevant to quantum information processing. Surprisingly, as will be demonstrated later, the ED can be determined through direct calculation, bypassing the challenges associated with the combinatorial nature of these states. Moreover, the resulting functional form can be interpreted in terms of the topological properties of the graphs associated with the various states in this class.

 The present work is organized as follows. In the section \ref{sec:gs}. \textit{Direct Graph States}, the structure of general undirected and directed graph states is defined. In the section \ref{sec:EGGC}. \textit{Entanglement in General Graph Configurations}, the Entanglement Distance is introduced and its application to the class of graph states is discussed. In \ref{sec:applications}. \textit{Applications}, the results obtained for the Entanglement Distance applied to various examples of states belonging to the class of graph states are reported. \ref{sec:concluding}. \textit{Concluding Remarks}  is the last section.

\section{Directed Graph States} \label{sec:gs}
At the base of the definition of a graph state is a graph, a collection of vertices and pairs of vertices connected by edges. Each graph is represented by a diagram in a plane, where the vertices are represented by points and the edges by arcs joining two vertices. The arcs are oriented, in the case of a directed graph, and not oriented in the case of an undirected graph. The most commonly implied graphs in this context are simple graphs, meaning they have no loops, at most one edge between any two vertices, and are undirected, meaning the edges do not have a specific direction. In the present work, we will consider directed graphs.

Mathematically, a directed simple graph $G(V,L)$ is a pair of sets $(V,L)$, where $V$ is the set of vertices, $V=\{1,\ldots,M\}$, $M\in \mathbb{N}^+$, and $L$ is the set of ordered couples of elements of $V$, identifying the set of oriented edges, $L=\{(a,b)|\ a,b\in V\}$. Let $M=| V|$ be the number of vertices and let $E=|L|$ be the number of edges in the graph. Furthermore, let $\Gamma$ be the oriented adjacency $M\times M$ matrix associated to the graph $G(V,L)$, with elements
\begin{equation}
    \Gamma_{ab} = \left\{
    \begin{array}{l}
        1  \ \textrm{if} \ (a,b) \in L   \\
        0 \ \textrm{otherwise}
    \end{array}
    \right.  \, .
\end{equation}
A directed graph state is a special pure multiparty quantum state of a distributed quantum system. It corresponds to a direct graph where each edge represents an Ising-like interaction between an ordered pairs of quantum spin systems or qubits. More precisely, one can provide the basic definitions for graph states with the notations introduced above. The graph state can be done using the stabilizer formalism \cite{hein_multi-party_2004}, nevertheless here we give their definition in terms of interaction patterns as follows.
To each vertex $a\in V$ is associated a qubit state $\ket{\phi}^{a}$, and to any ordered pair $(a,b)$ of connected vertices is associated a nonlocal operator
\begin{equation}
\label{5}
{U}_{ab}=\Pi^{(a)}_0\mathbb{I}^{(b)}+\Pi^{(a)}_1\bar{U}^{(b)} \, ,
\end{equation}
where $\Pi^{a}_0 (\Pi^{a}_1)$ is the projector of the subspace of the $a$-th qubit onto the state $\ket{0}^{a} (\ket{1}^{a})$, $\mathbb{I}^{b}$ and $\bar{U}^b$ are respectively the identity matrix and a generic $U(2)$ operator acting on the subspace of the $b$-th qubit. Here, by $\ket{0}$ and $\ket{1}$ we denote the computational basis, i.e., the eigenstates of the $\sigma_z$ operator with eigenvalues $+1$ and $-1$ respectively. The matrix $\bar{U}^{b}$ determines the interaction strength, which is assumed to be the same for all the connected pairs. The graph state associated to a graph $G(V,L)$ is
\begin{equation}
    \ket{G}=\prod_{(a,b) \in L}{U}_{ab}\ket{\phi}^{\otimes M} \, .
\end{equation}
where we assumed that each qubit starts from the same single-qubit state, represented by the ket $\ket{\phi}$. In this framework, a completely empty graph corresponds to the state $\ket{\phi}^{\otimes M}$. This assumption is not only the simplest and generally adopted basis for further development but is also crucial for performing our topological analysis of the quantum graph via entanglement measures. 

\section{Entanglement in General Graph Configurations} \label{sec:EGGC}

\subsection{General Framework}

In the present study, we adopt the Entanglement Distance \cite{cocchiarella_entanglement_2020,vesperini_entanglement_2023, Vesperini_2023, Vesperini_2024} as a measure of entanglement. The ED is derived from the Fubini-Study metric associated with the projective Hilbert space. The ED per qubit is
\begin{equation}\label{single-qubit_ED}
E(\ket{G})=1-\dfrac{1}{M}\sum_{i=1}^{M}||\bra{G}\boldsymbol{\sigma}^{(i)}\ket{G}||^2 \, ,
\end{equation}
where $\boldsymbol{\sigma}^{(i)}=(\sigma^{(i)}_x,\sigma^{(i)}_y,\sigma^{(i)}_z)$ is the vector of the Pauli matrices operating on the qubit $i$.
The ED equals $M$ if $\ket{G}$ is maximally entangled, and $0$ if it is fully separable.

The interaction pattern of a graph state is completely specified by a simple graph G, iff {\it i)} there is no ordering of the edges, namely all two–particle unitary operators commutate, that is $[{U}_{ab}, {U}_{bc}]=0$, $\forall a,b,c \in V$; {\it ii)} all qubits interact through the same two–particle unitary operator $U_{ab}=U$, $\forall (a,b)\in L$.

These conditions can be satisfied by significantly restricting the single-qubit operator $\bar{U}$ to the form
\begin{equation}
\bar{U} = e^{-i\psi}
\begin{pmatrix}
e^{i\theta} & 0  \\
0 & {e^{-i\theta}}
\end{pmatrix}
\label{mtxU}
\end{equation}
with $\psi,\theta \in \mathbb{R}$.

It is worth emphasizing that condition {\it i)} --namely, that all two-qubit operators commute-- while restrictive, is consistent with the standard literature on graph states. This requirement is necessary in order to derive a state directly from a graph, where edges are not endowed with any intrinsic ordering.
For this reason, the results of the present work hold under assumption {\it i)}.

\subsection{Initial State}

To determine the initial state $|\phi\rangle^{\otimes M}$ that allows one to generate a maximally entangled network, we consider two general vertices $a,b\in V$, connected by an arc, and derive the single-vertex reduced density matrices 
$\rho_a = tr_b [\rho_{ab}]$ , and $\rho_b = tr_a [\rho_{ab}]$, where $\rho_{ab} = \ket{\psi_{ab}}\bra{\psi_{ab}}$ is the density matrix of the entangled state 
\begin{equation}
\ket{\psi_{ab}}=U_{ab} \ket{\phi}^{\otimes 2} \, .
\label{psiab}
\end{equation}
Therefore, the initial state is determined by the constraint that the distance between the single-vertex density matrix $\rho_a$ ($\rho_b$), and the single-qubit density matrix $\rho_M =\mathbb{I}/2$ corresponding to the maximum entanglement, is zero.

The Hilbert-Schmidt distance $D_{HS}$ gives the distance between two square matrices $\rho_1$ and $\rho_2$ as
\begin{equation}
D_{HS}(\rho_1,\rho_2) = \sqrt{\dfrac{1}{2} \mathrm{tr} \big[(\rho_1-\rho_2)^\dagger (\rho_1-\rho_2)\big]} \, .
\end{equation}
With this, one can derive the distance between the single-vertex density matrix $\rho_a$ (or equivalently $\rho_b$) and $\rho_M$.
We have
\begin{align*}
\rho_{ab}=&\ket{\psi_{ab}}\bra{\psi_{ab}}=\Pi_0\rho\,\Pi_0\otimes\rho +\Pi_1\rho\,\Pi_1\otimes \bar{U} \rho \bar{U}^\dagger + \\ &+\bigl(\Pi_0\rho\,\Pi_1\otimes  \rho \bar{U}^{\dagger}+\text{h.c.}\bigr)
\end{align*}
where $\rho=\ket{\phi}\bra{\phi}$. In the case of a general state $\ket{\phi}=\alpha_0\ket{0}+\alpha_1\ket{1}$ with $|\alpha_0|^2 + |\alpha_1|^2 =1$, the square of the Hilbert-Schmidt distance is
\begin{align*}
& D^2_{HS}\left(\rho_\gamma,\rho_M\right)= \dfrac{1}{4}\!-\!2 p^2\!+\!4 p^3\!-\!2 p^4\!+\!2 (1\!-\!p)^2p^2 \cos (2\theta) \, , 
\end{align*}
where $\gamma =a,b$, and $p=|\alpha_1|^2$.

\begin{figure}[h]
    \centering
    \includegraphics[width=\columnwidth]{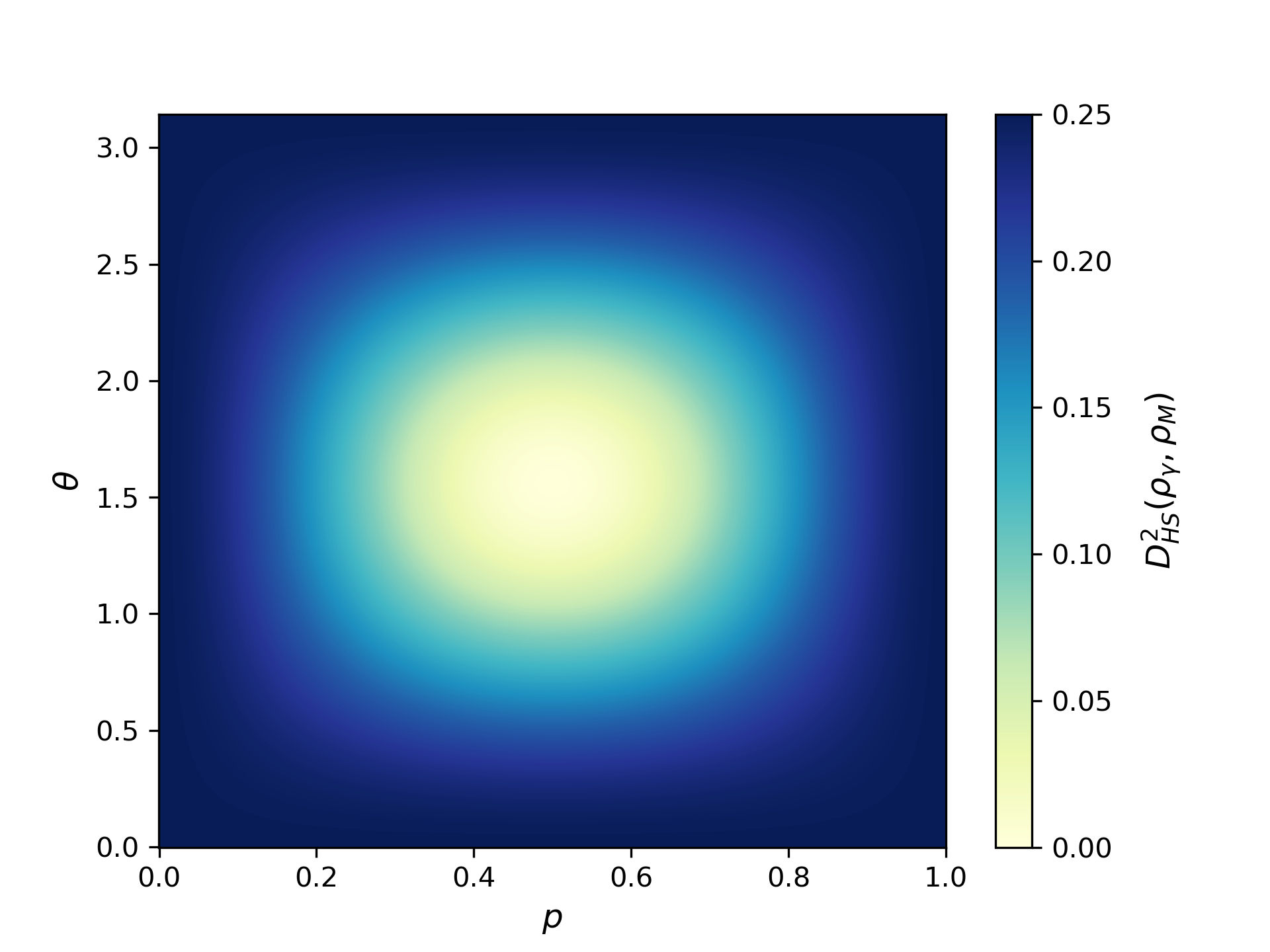}
    \caption{This figure reports the color map of $D^2_{HS} (\rho_\gamma,\rho_M)$, for $\gamma=a,b$, as a function of $p$ and $\theta$.\label{fig01}}

\end{figure}

Fig. (\ref{fig01}) shows the color map of $D^2_{HS} (\rho_\gamma,\rho_M)$, for $\gamma =a,b$, as a function of $p$ and $\theta$.
Fig. (\ref{fig01}) clearly shows that the distance between $\rho_\gamma$, for $\gamma =a,b$ and $\mathbb{I}/2$ is zero if and only if $p=1/2$ and $\theta = \pi/2$.

\begin{figure}[h]
    \centering
    \includegraphics[width=\columnwidth]{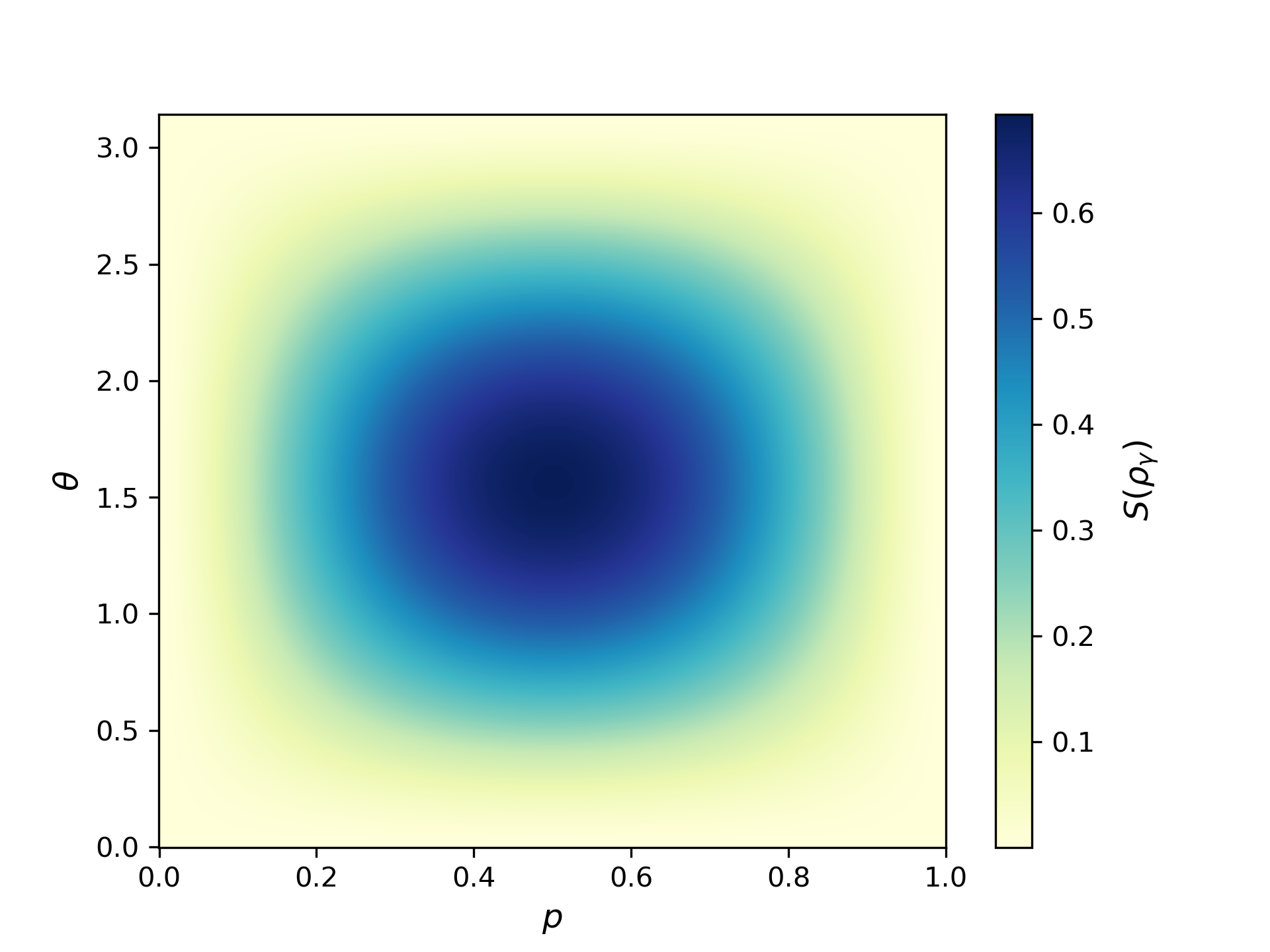}
    \caption{von Neumann entropy $S(\rho_\gamma)$, for $\gamma = a,b$, as a function of $p$ and $\theta$.\label{fig02}}
    
\end{figure}

Alternatively, the initial state $\ket{\phi}^{\otimes M}$ that allows one to generate maximally entangled states can also be derived through the analysis of entanglement, either by the von Neumann entropy of the reduced density matrices $\rho_\gamma$, $\gamma=a,b$, which is an entanglement measure for bipartite systems, or by calculating the ED of the pure state $\ket{\psi_{ab}}$. By directly calculating the eigenvalues of $\rho_\gamma$, for $\gamma=a,b$, one obtains
\begin{equation}
\lambda_j = \frac{1}{2}\left[(-1)^j  \sqrt{1-16 p^2(1-p)^2 \sin^2(\theta)} + 1\right] \, ,
\end{equation}
for $j=1,2$, and then the von Neumann entropy $S(\rho_{\gamma})=-\mathrm{tr}[\rho_\gamma \ln \rho_\gamma] = - \sum_{j=1}^2 \lambda_j \ln [\lambda_j]$, for $\gamma=a,b$.
Fig. (\ref{fig02}) shows the color map of $S(\rho_\gamma)$, for $\gamma =a,b$, as a function of $p$ and $\theta$.
A direct calculation of the ED (\ref{single-qubit_ED}) for the state \eqref{psiab} yields
\begin{equation}
E(\ket{\psi_{ab}})=16p^2(1-p)^2\sin^2\theta \, .
\end{equation}
Both the von Neumann entropy $S$ and the entanglement distance $ E(\ket{\psi_{ab}})$ reach their maximum at the point $(p,\theta)=(\frac{1}{2},\frac{\pi}{2})$. Furthermore, along the line $p=1/2$, the eigenvalues of $\rho_\gamma$, $\gamma=a,b$, are $\lambda_{1,2}=\frac{1}{2}\pm\frac{1}{2}|\cos\theta|$, and the von Neumann entropy results
\begin{equation}
S=\log\bigg(\frac{2}{|\sin\theta|}\bigg)+\frac{1}{2}|\cos\theta|\log\bigg(\dfrac{1-|\cos\theta|}{1+|\cos\theta|}\bigg) \, .
\end{equation}

From the calculations reported above, it is clear that the entanglement properties of a graph state do not depend on the phase of $\alpha_0$ and $\alpha_1$ in the initial state $\ket{\phi}$, but rather on their modulus. Additionally, a maximally entangled network is obtained only when $|\alpha_0|=|\alpha_1|= 1/\sqrt{2}$. Therefore, from now on, we will adopt the initial state $\ket{\phi} = (\ket{0} + \ket{1})/\sqrt{2}$.

\subsection{Entanglement Distance Computation}

To compute the ED \eqref{single-qubit_ED} for a general graph $G(V,L)$, one has to calculate the value $||\bra{G}\boldsymbol{\sigma}^{(i)}\ket{G}||^2$ for each $i \in V$. Let $\Gamma_\rightarrow(i)$ ($\Gamma_\leftarrow(i)$) denote the set of vertex labels connected to $i$ by an outgoing (incoming) link, 
\begin{align}
\Gamma_\rightarrow(i) = \{j\in V| (i,j)\in L  \} \, , \\
\Gamma_\leftarrow(i) = \{j\in V|  (j,i) \in L \} \, .
\end{align}
The set $\Gamma(i)=\Gamma_{\rightarrow}(i)\cup\Gamma_{\leftarrow}(i)$ thus is the set of vertices connected to $i$ by an edge, and $d(i):=|\Gamma(i)|$ the degree of the vertex $i$. 

Remarkably, the numerical value of the contribution to the ED from a generic vertex $i$ depends on $d(i)$ rather than on the individual values of $|\Gamma_\rightarrow (i)|$ and $|\Gamma_\leftarrow (i)|$, as demonstrated in \cite{desimone2025entanglementdirectedgraphstates}.

Furthermore, the numerical value of the contribution to the ED arising from the $i$-th vertex, as well as the total ED, remains invariant under vertex renumbering. Thus, we expect that the ED depends only on the local graph topology, specifically the coordination number of each vertex. Therefore,  set $E(\theta;\{d(i)\}):=E(\ket{G})$. In Ref. \cite{desimone2025entanglementdirectedgraphstates}, we have shown that the ED per qubit for a general graph is 
\begin{equation}
\label{final}
E(\theta;\{d(i)\})=1-\dfrac{1}{M}\sum_{i\in V} [\cos ({\theta})]^{2d(i)} \, .
\end{equation}
Notably, the ED is independent of the parameter $\psi$ in $\bar{U}$. Since entanglement in the present case, does not depend on the orientation of the edges, from now on, we will describe the graph topology purely in terms of its connectivity, without distinguishing between outgoing or incoming links. 
A more general derivation, extending the arguments in Ref. \cite{desimone2025entanglementdirectedgraphstates}, is presented in Appendix \ref{appendix}. There, we consider a generic state $|\tilde{\phi}\rangle = \alpha_0 \ket{0} + \alpha_1 \ket{1}$, with $|\alpha_0|^2 + |\alpha_1|^2 = 1$, and show that the Entanglement Distance still depends solely on the degree distribution of the graph, with Eq. \eqref{final} arising as a particular case of this general result.

The formula \eqref{final}                         
        can be expressed in terms of the degree distribution of the vertices. Let $n_k$ denote the number of vertices with a degree of connection equal to $k$.  Then, we have
\begin{equation}
\sum_{i \in V} [\cos ({\theta})]^{2d(i)} = \sum_{k=0}^{|L|} n_k [\cos ({\theta})]^{2k} \, ,
\end{equation}
and the ED per qubit becomes
\begin{equation}
\label{final2}
    E\left(\theta;\{n_k\}\right)=1-\dfrac{1}{M} \sum_{k=0}^{|L|} n_k[\cos(\theta)]^{2k}\, .
\end{equation}

In addition, in the following examples, we consider graphs with a somewhat regular structure, which allows us to view them as the result of a sequence of elementary operations on subgraphs. These elementary operations consist of adding vertices and appropriate links to a subgraph.

The general scheme is as follows.
For each graph $G(V,L)$ in the examples, we identify an integer $K$ and consider a family of subgraphs $G_k(V_k,L_k)$, for $k=1,\ldots,K$, such that
\begin{equation}
G_1(V_1,L_1) \subset G_2(V_2,L_2) \subset \cdots G_K(V_K,L_K) =G(V,L) \, ,
\end{equation}
where the symbol $\subset$ indicates that each subgraph is a proper subgraph of the following one.
Furthermore, we have
\begin{equation}
    G_{k+1}(V_{k+1},L_{k+1}) = G_k(V_k,L_k) + g_k(v_k,l_k)
\end{equation}
here, the symbol $+$ denotes the joining of the two subgraphs $G_k(V_k,L_k)$ and $g_k(v_k,l_k)$, meaning they are joined together by additional edges $X_k$. The set of vertices $v_k$ identifies a new layer $C_k$ in $G_{k+1}(V_{k+1},L_{k+1})$ such that
$v_k\cap V_k=\emptyset$, and $V_{k+1}=V_k\cup v_k$. The sets of edges $l_k$ and $L_k$ are disjoint, and during the joining operation, the additional edges $X_k$ are added, so that $L_{k+1}=L_k \cup l_k \cup X_k$.


\section{Applications} \label{sec:applications}
\subsection{Variant of Young-Fibonacci graph\label{application A}}
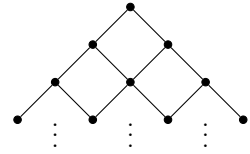
\begin{figure}[h]
\begin{center}
\begin{tikzpicture}
  \node[draw, fill=black, circle, inner sep=1pt] (A) at (0, 0) {};
  \node[draw, fill=black, circle, inner sep=1pt] (B) at (-0.5, -0.5) {};
  \node[draw, fill=black, circle, inner sep=1pt] (C) at (0.5, -0.5) {};
  \node[draw, fill=black, circle, inner sep=1pt] (D) at (-1, -1) {};
  \node[draw, fill=black, circle, inner sep=1pt] (E) at (0, -1) {};
  \node[draw, fill=black, circle, inner sep=1pt] (F) at (1, -1) {};
  \node[draw, fill=black, circle, inner sep=1pt] (G) at (-1.5, -1.5) {};
  \node[draw, fill=black, circle, inner sep=1pt] (H) at (-0.5, -1.5) {};
  \node[draw, fill=black, circle, inner sep=1pt] (I) at (0.5, -1.5) {};
  \node[draw, fill=black, circle, inner sep=1pt] (L) at (1.5, -1.5) {};
  \draw (A) -- (B);
  \draw (A) -- (C);
  \draw (B) -- (D);
  \draw (B) -- (E);
  \draw (C) -- (E);
  \draw (C) -- (F);
  \draw (D) -- (G);
  \draw (D) -- (H);
  \draw (E) -- (H);
  \draw (E) -- (I);
  \draw (F) -- (I);
  \draw (F) -- (L);
  \node at (-1, -1.6) {\vdots};
  \node at (0, -1.6) {\vdots};
  \node at (1, -1.6) {\vdots};
\end{tikzpicture}
\caption{Graph topology for the ED per qubit (\ref{appA}).\label{FigA}}
\end{center}
\end{figure}

The topology of this type of graph is shown in Fig. (\ref{FigA}). Let $C_i$ be the $i$-th layer, with $C_1$ being the top layer and $C_N$ the bottom layer, where $N$ is the total number of layers. Denoting by $v_i$ the set of the vertices in the $C_i$-th layer, the total number of vertices is $M=\sum_{i=1}^N i = N(N+1)/2$, where $i=|v_i|$ is the number of vertices in the $i$-th layer. 
Let $n^{(i)}_k$ denote the number of vertices with a degree of connection equal to k, in the $i$-th layer. It is straightforward to recognize that the degree distribution is: \emph{i)} for $i\!=\!1$, $n_k^{(i)}=\delta_{k,2}$; \emph{ii)} for $i\!=\!2,\dots,N\!-\!1$, $n_k^{(i)}=2\delta_{k,3}+(i\!-\!2)\delta_{k,4}$; \emph{iii)} for $i\!=\!N$, $n_k^{(i)}=2\delta_{k,1}+(N\!-\!2)\delta_{k,2}$, where $\delta_{j,k}$ denotes the $(j,k)$-th element of the Kronecker delta. Therefore, the function $N(d)$ is $N(d)=2\delta_{d,1}+(N\!-\!1)\delta_{d,2}+2(N\!-\!2)\delta_{d,3}+\frac{(N\!-\!2)(N\!-\!3)}{2}\delta_{d,4}$ and the ED per qubit (\ref{final2}) is
\begin{equation}
\label{appA}
    \begin{split}
        E(\theta; N)=&1-\frac{\cos^2\theta}{N(N+1)}\Big(4+2(N\!-\!1)\cos^2\theta+\\
        &+4(N\!-\!2)\cos^4\theta+(N\!-\!2)(N\!-\!3)\cos^6\theta\Big)\, .
    \end{split}
\end{equation}

\begin{figure}[h]
    \includegraphics[width=\columnwidth]{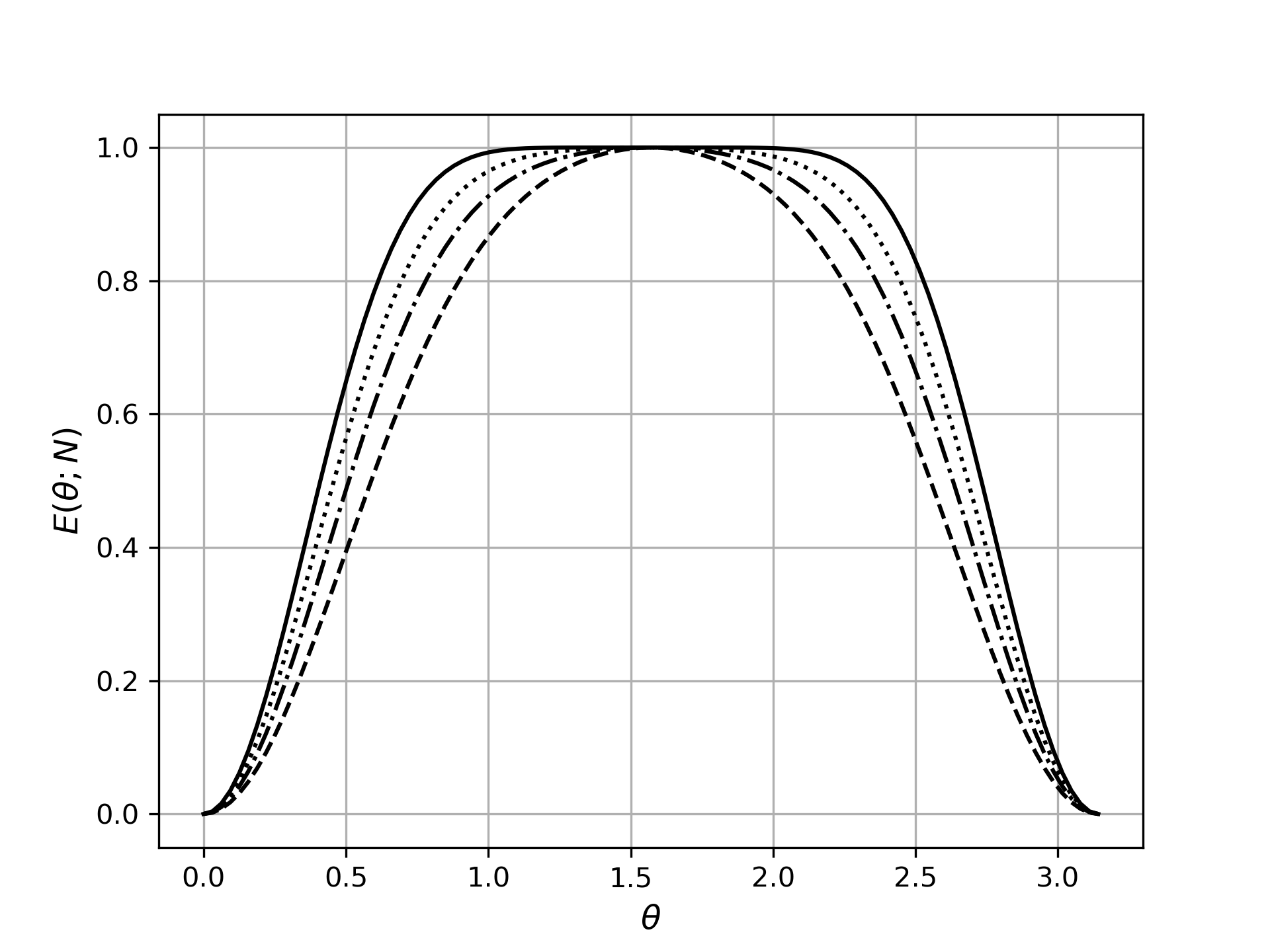}
    \caption{The figure reports the entanglement $E(\theta;N)$ for $N\!=\!3$ (dashed line), $N\!=\!5$ (dot-dashed line), $N\!=\!10$ (dotted line) and $N\!=\!+\infty$ (continuous line).\label{plotA}}
\end{figure}

We see that, in the limit $N\to+\infty$, the entanglement per qubit asymptotically approaches the bound
\begin{equation}
\label{appAinfty}
E(\theta;+\infty)=1-\cos^8{\theta}\, .
\end{equation}
This suggests that the predominant contribution comes from the internal vertices. Fig. (\ref{plotA}) shows the plots of this function for various values of $N$.
\subsection{Deep Feed Forward Neural Network}
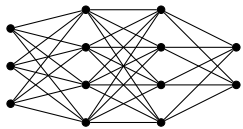
\begin{figure}[h]
\begin{center}
\begin{tikzpicture}
  \node[draw, fill=black, circle, inner sep=1pt] (1) at (0, 0) {};
  \node[draw, fill=black, circle, inner sep=1pt] (2) at (0, 0.5) {};
  \node[draw, fill=black, circle, inner sep=1pt] (3) at (0, -0.5) {};

  \node[draw, fill=black, circle, inner sep=1pt] (4) at (1, 0.75) {};
  \node[draw, fill=black, circle, inner sep=1pt] (5) at (1, 0.25) {};
  \node[draw, fill=black, circle, inner sep=1pt] (6) at (1, -0.25) {};
  \node[draw, fill=black, circle, inner sep=1pt] (7) at (1, -0.75) {};

  \node[draw, fill=black, circle, inner sep=1pt] (8) at (2, 0.75) {};
  \node[draw, fill=black, circle, inner sep=1pt] (9) at (2, 0.25) {};
  \node[draw, fill=black, circle, inner sep=1pt] (10) at (2, -0.25) {};
  \node[draw, fill=black, circle, inner sep=1pt] (11) at (2, -0.75) {};

  \node[draw, fill=black, circle, inner sep=1pt] (12) at (3, 0.25) {};
  \node[draw, fill=black, circle, inner sep=1pt] (13) at (3, -0.25) {};
  
  \draw (1) -- (4);
  \draw (1) -- (5);
  \draw (1) -- (6);
  \draw (1) -- (7);
  \draw (2) -- (4);
  \draw (2) -- (5);
  \draw (2) -- (6);
  \draw (2) -- (7);
  \draw (3) -- (4);
  \draw (3) -- (5);
  \draw (3) -- (6);
  \draw (3) -- (7);
  
  \draw (4) -- (8);
  \draw (4) -- (9);
  \draw (4) -- (10);
  \draw (4) -- (11);
  \draw (5) -- (8);
  \draw (5) -- (9);
  \draw (5) -- (10);
  \draw (5) -- (11);
  \draw (6) -- (8);
  \draw (6) -- (9);
  \draw (6) -- (10);
  \draw (6) -- (11);
  \draw (7) -- (8);
  \draw (7) -- (9);
  \draw (7) -- (10);
  \draw (7) -- (11);

  \draw (8) -- (12);
  \draw (8) -- (13);
  \draw (9) -- (12);
  \draw (9) -- (13);
  \draw (10) -- (12);
  \draw (10) -- (13);
  \draw (11) -- (12);
  \draw (11) -- (13);

\end{tikzpicture}
\caption{An example of a recurrent neural network, with $M_1=3$, $M_2=4$, $M_3=4$, $M_4=2$.\label{network}}
\end{center}
\end{figure}

The topology of this graph consists of an input layer $C_1$, an output layer $C_N$, and multiple hidden layers $C_i$ for $i\!=\!2,\dots,N\!-\!1$. Let $v_i$ be the set of neurons in the $i$-th layer, and let $M_i=|v_i|$ denote the number of neurons in that layer. An example of this graph is shown in Fig. (\ref{network}). The distribution of the degrees is as follows; \emph{i)} for $i\!=\!1$, $n_k^{(i)}=M_1\delta_{k,M_2}$; \emph{ii)} for $i\!=\!2,\dots,N\!-\!1$, $n_k^{(i)}=M_i\delta_{k,M_{i+1}+M_{i-1}}$; \emph{iii)} for $i\!=\!N$, $n_k^{(i)}=M_N\delta_{k,M_{N-1}}$.
The function $N(d)$ is $N(d)=M_1\delta_{d,M_2}+\sum_{i=2}^{N-1}M_i\delta_{d,M_{i+1}+M_{i-1}}+M_N\delta_{d,M_{N-1}}$ and the entanglement distance (\ref{final2}) becomes
\begin{equation}
\begin{split}
    E\left(\theta;\{M_i\}\right)=&1-\frac{1}{M}\Big(M_1[\cos(\theta)]^{2M_2}+M_N[\cos(\theta)]^{2M_N}+\\+ &\sum_{i=2}^{N-1}M_i[\cos(\theta)]^{2(M_{i+1}+M_{i-1})}\Big)\,,
\end{split}
\end{equation}
where $M=\sum_{i=1}^N M_i$ represents the total number of vertices.
\subsection{Full Binary Tree}
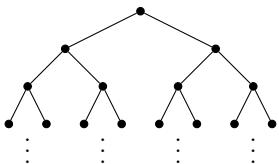
\begin{figure}[h]
\begin{center}
    \begin{tikzpicture}
         \node[draw, fill=black, circle, inner sep=1pt] (1) at (0, 0) {};
         
         \node[draw, fill=black, circle, inner sep=1pt] (2) at (-1, -0.5) {};
         \node[draw, fill=black, circle, inner sep=1pt] (3) at (1, -0.5) {};
         
         \node[draw, fill=black, circle, inner sep=1pt] (4) at (-1.5, -1.) {};
         \node[draw, fill=black, circle, inner sep=1pt] (5) at (-0.5, -1.) {};
         \node[draw, fill=black, circle, inner sep=1pt] (6) at (0.5, -1.) {};
         \node[draw, fill=black, circle, inner sep=1pt] (7) at (1.5, -1.) {};
         
         \node[draw, fill=black, circle, inner sep=1pt] (8) at (-1.75, -1.5) {};
         \node[draw, fill=black, circle, inner sep=1pt] (9) at (-1.25, -1.5) {};
         \node[draw, fill=black, circle, inner sep=1pt] (10) at (-0.75, -1.5) {};
         \node[draw, fill=black, circle, inner sep=1pt] (11) at (-0.25, -1.5) {};
         \node[draw, fill=black, circle, inner sep=1pt] (12) at (0.25, -1.5) {};
         \node[draw, fill=black, circle, inner sep=1pt] (13) at (0.75, -1.5) {};
         \node[draw, fill=black, circle, inner sep=1pt] (14) at (1.25, -1.5) {};
         \node[draw, fill=black, circle, inner sep=1pt] (15) at (1.75, -1.5) {};
  \node at (-1.5, -1.75) {\vdots};
  \node at (-0.5, -1.75) {\vdots};
  \node at (0.5, -1.75) {\vdots};
  \node at (1.5, -1.75) {\vdots};
         \draw (1)--(2);
         \draw (1)--(3);
         \draw (2) -- (4);
         \draw (2) -- (5);
         \draw (3) --(6);
         \draw(3)--(7);
         \draw (4)--(8);
         \draw(4)--(9);
         \draw(5)--(10);
         \draw(5)--(11);
         \draw(6)--(12);
         \draw(6)--(13);
         \draw(7)--(14);
         \draw(7)--(15);
    \end{tikzpicture}
    \caption{The figure reports the graph for the ED per qubit in (\ref{binarytree}).\label{FigC}}
    \end{center}
\end{figure}
\begin{figure}[h]
    \includegraphics[width=\columnwidth]{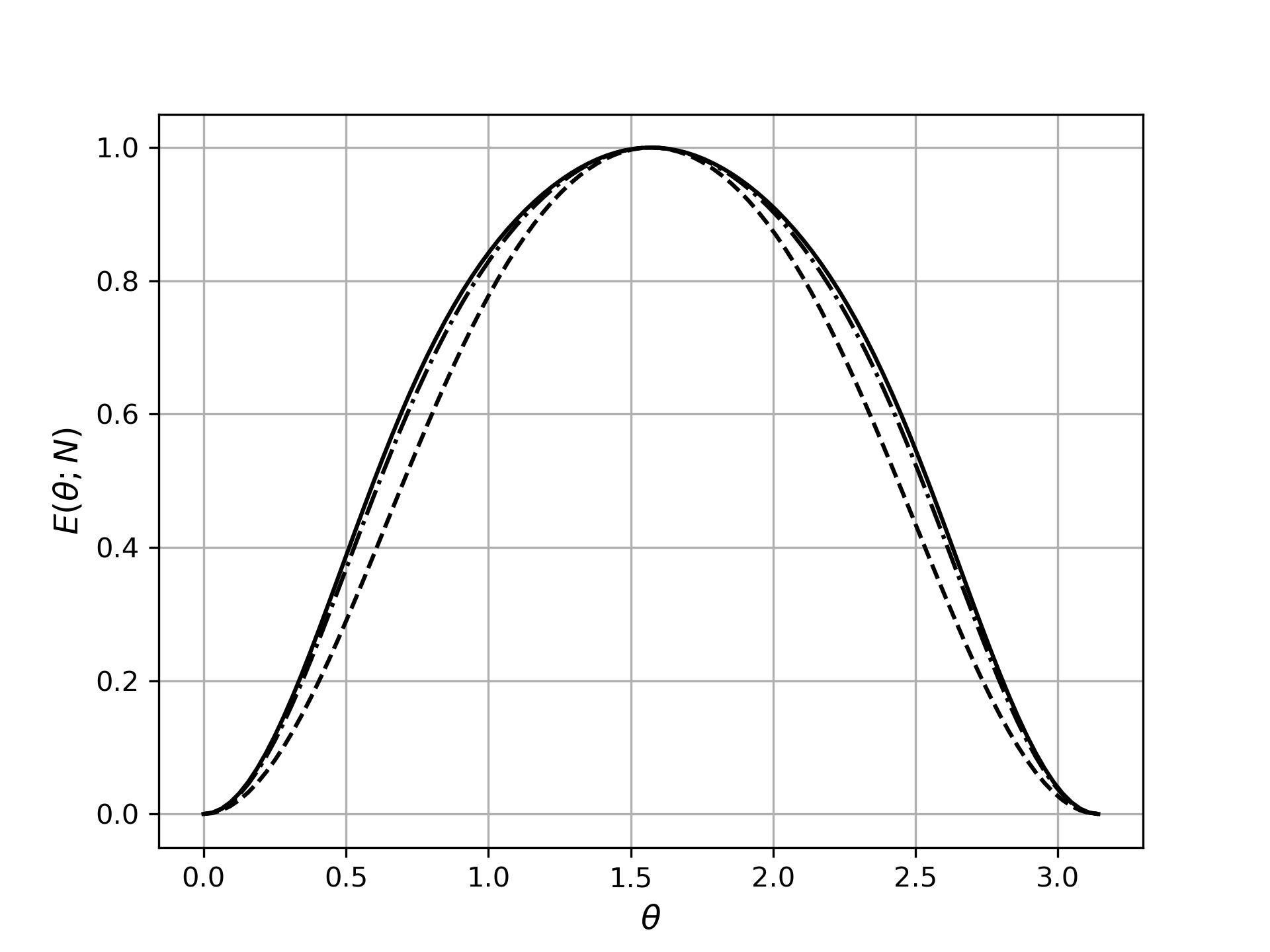}
    \caption{The figure reports the ED per qubit for $N=2$ (dashed line), $N=4$ (dot-dashed line) and $N=+\infty$ (continuous line).\label{plotC}}
\end{figure}

The structure of this graph is shown in Fig. (\ref{FigC}). The division into subgraphs is similar to the case in (\ref{application A}), where now the number of vertices in the $i$-th layer is $|v_i|=2^{i\!-\!1}$ and the total number of vertices is $M=2^N\!-\!1$. For the distribution of the degrees, we have; \emph{i)} for $i\!=\!1$, $n_k^{(i)}=\delta_{k,2}$; \emph{ii)} for $i\!=\!2,\dots,N\!-\!1$, $n_k^{(i)}=2^{i-1}\delta_{k,3}$; \emph{iii)} for $i\!=\!N$, $n_k^{(i)}=2^{N-1}\delta_{k,1}$. Therefore, the degrees distribution is $N(d)=2^{N-1}\delta_{d,1}+\delta_{d,2}+2(2^{N-2}-1)\delta_{d,3}$ and the entanglement per qubit (\ref{final2}) becomes
\begin{equation}
\label{binarytree}
    E(\theta;N)=1-\frac{\cos^2\theta}{2^N-1}\Big(2^{N-1}\!+\!\cos^2\theta+(2^{N-1}\!-\!2)\cos^4\theta\Big)\, .
\end{equation}
It is notable that, as the number of layers approaches to infinity, the entanglement asymptotically approaches
\begin{equation}
    E(\theta;+\infty)=1-\frac{\cos^2\theta}{2}\Big(1+\cos^4\theta\Big)\, ,
\end{equation}
i.e., the predominant contribution comes from internal and last vertices. In Fig. (\ref{plotC}) we plot $E(\theta;N)$ for various values of $N$. 

\subsection{Linear Bridged Cycle Graph}
\begin{figure}
\begin{center}
\begin{tikzpicture}
\coordinate (a) at (0,0);
\coordinate (b) at ({3*cos(-90)/4}, {3*sin(-90)/4});
\coordinate (bf) at ({3*cos(90)/4}, {3*sin(90)/4});

\coordinate (aa) at (2,0);
\coordinate (c) at ({3*cos(-45)/4+2}, {3*sin(-45)/4});
\coordinate (cf) at ({3*cos(225)/4+2}, {3*sin(225)/2});
\coordinate (c1) at ({3*cos(45)/4+2}, {3*sin(45)/4});
\coordinate (cf1) at ({3*cos(135)/4+2}, {3*sin(135)/2});

\coordinate (aaa) at (4,0);
\coordinate (d) at ({3*cos(-45)/4+4}, {3*sin(-45)/4});
\coordinate (df) at ({3*cos(225)/4+4}, {3*sin(225)/4});
\coordinate (d1) at ({3*cos(45)/4+4}, {3*sin(45)/4});
\coordinate (df1) at ({3*cos(135)/4+4}, {3*sin(135)/4});

\coordinate (aaaa) at (6,0);
\coordinate (e) at ({3*cos(270)/4+6}, {3*sin(270)/4});
\coordinate (ef) at ({3*cos(90)/4+6}, {3*sin(90)/4});

  \node[draw, fill=black, circle, inner sep=1pt, minimum size=1pt] (1) at ({3*cos(-45)/4}, {3*sin(-45)/4}) {};
  \node[draw, fill=black, circle, inner sep=1pt, minimum size=1pt] (2) at ({3*cos(0)/4}, {3*sin(0)/4}) {};
  \node[draw, fill=black, circle, inner sep=1pt, minimum size=1pt] (3) at ({3*cos(45)/4}, {3*sin(45)/4}) {};
  \node[draw, fill=black, circle, inner sep=1pt, minimum size=1pt] (4) at ({3*cos(90)/4}, {3*sin(90)/4}) {};
  \node[draw, fill=black, circle, inner sep=1pt, minimum size=1pt] (101) at ({3*cos(-90)/4}, {3*sin(-90)/4}) {};

  \node[draw, fill=black, circle, inner sep=1pt, minimum size=1pt] (6) at ({3*cos(-45)/4+2}, {3*sin(-45)/4}) {};
  \node[draw, fill=black, circle, inner sep=1pt, minimum size=1pt] (7) at ({3*cos(0)/4+2}, {3*sin(0)/4}) {};
  \node[draw, fill=black, circle, inner sep=1pt, minimum size=1pt] (8) at ({3*cos(45)/4+2}, {3*sin(45)/4}) {};
  
  \node[draw, fill=black, circle, inner sep=1pt, minimum size=1pt] (10) at ({3*cos(135)/4+2}, {3*sin(135)/4}) {};
  \node[draw, fill=black, circle, inner sep=1pt, minimum size=1pt] (11) at ({3*cos(180)/4+2}, {3*sin(180)/4}) {};
  \node[draw, fill=black, circle, inner sep=1pt, minimum size=1pt] (102) at ({3*cos(225)/4+2}, {3*sin(225)/4}) {};

  \node[draw, fill=black, circle, inner sep=1pt, minimum size=1pt] (12) at ({3*cos(-45)/4+4}, {3*sin(-45)/4}) {};
  \node[draw, fill=black, circle, inner sep=1pt, minimum size=1pt] (13) at ({3*cos(0)/4+4}, {3*sin(0)/4}) {};
  \node[draw, fill=black, circle, inner sep=1pt, minimum size=1pt] (14) at ({3*cos(45)/4+4}, {3*sin(45)/4}) {};
  
  \node[draw, fill=black, circle, inner sep=1pt, minimum size=1pt] (16) at ({3*cos(135)/4+4}, {3*sin(135)/4}) {};
  \node[draw, fill=black, circle, inner sep=1pt, minimum size=1pt] (17) at ({3*cos(180)/4+4}, {3*sin(180)/4}) {};
  \node[draw, fill=black, circle, inner sep=1pt, minimum size=1pt] (100) at ({3*cos(225)/4+4}, {3*sin(225)/4}) {};

  \node[draw, fill=black, circle, inner sep=1pt, minimum size=1pt] (18) at ({3*cos(90)/4+6}, {3*sin(90)/4}) {};
  \node[draw, fill=black, circle, inner sep=1pt, minimum size=1pt] (19) at ({3*cos(135)/4+6}, {3*sin(135)/4}) {};
  \node[draw, fill=black, circle, inner sep=1pt, minimum size=1pt] (20) at ({3*cos(180)/4+6}, {3*sin(180)/4}) {};
  \node[draw, fill=black, circle, inner sep=1pt, minimum size=1pt] (21) at ({3*cos(225)/4+6}, {3*sin(225)/4}) {};
  \node[draw, fill=black, circle, inner sep=1pt, minimum size=1pt] (22) at ({3*cos(270)/4+6}, {3*sin(270)/4}) {};

  \path[-] (1) edge node {} (2);
  \path[-] (2) edge node {} (3);
  \path[-] (3) edge node {} (4);
  \path[-] (101) edge node {} (1);

  \path[-] (6) edge node {} (7);
  \path[-] (7) edge node {} (8);
  
  \path[-] (10) edge node {} (11);
  \path[-] (102) edge node {} (11);
  
  \path[-] (12) edge node {} (13);
  \path[-] (13) edge node {} (14);
  
  \path[-] (16) edge node {} (17);
  \path[-] (100) edge node {} (17);

  \path[-] (18) edge node {} (19);
  \path[-] (19) edge node {} (20);
  \path[-] (20) edge node {} (21);
  \path[-] (21) edge node {} (22);

  \path[-] (2) edge node {} (11);
  \path[-] (7) edge node {} (17);
  
  \node at (5, 0) {$\cdot\mkern-4mu\cdot\mkern-4mu\cdot$};
  
  \pic[draw, dashed, angle radius=0.75cm,angle eccentricity=0] {angle=bf--a--b};
  \pic[draw, dashed, angle radius=0.75cm,angle eccentricity=0] {angle=cf--aa--c};
  \pic[draw, dashed, angle radius=0.75cm,angle eccentricity=0] {angle=c1--aa--cf1};
  \pic[draw, dashed, angle radius=0.75cm,angle eccentricity=0] {angle=df--aaa--d};
  \pic[draw, dashed, angle radius=0.75cm,angle eccentricity=0] {angle=d1--aaa--df1};
  \pic[draw, dashed, angle radius=0.75cm,angle eccentricity=0] {angle=e--aaaa--ef};
\end{tikzpicture}
\caption{The figure shows multiple cycle graphs, each connected by a single edge between distinct vertices. \label{FigD}}
\end{center}
\end{figure}
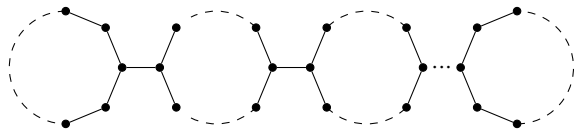
The design of this graph is shown in Fig. (\ref{FigD}). Let $C_i$ be the $i$-th circle, with $C_1$ being the leftmost circle and $C_N$ the rightmost. We denote by $v_i$ the set of the vertices in the $C_i$-th circle, and by $M=\sum_{i=1}^N M_i$ the total number of vertices, where $M_i=|v_i|$ represents the number of vertices in the $i$-th circle. The distribution of the degrees is as follows; \emph{i)} for $i\!=\!1$, $n_k^{(i)}=(M_1\!-\!1)\delta_{2,k}+\delta_{3,k}$; \emph{ii)} for $i\!=\!2,\dots,N\!-\!1$, $n_k^{(i)}=(M_i\!-\!2)\delta_{2,k}+2\delta_{3,k}$; \emph{iii)} for $i\!=\!N$, $n_k^{(i)}=(M_N\!-\!1)\delta_{2,k}+\delta_{3,k}$. Therefore, the number $N(d)$ is $N(d)=2(N\!-\!1)\delta_{3,d}+(M-2N+2)\delta_{2,d}$ and the ED per qubit (\ref{final}) is
\begin{equation}
    E(\theta;M,N)=1-\frac{\cos^4 \theta}{M}\Big(M\!-\!2(N\!-\!1)\sin^2\theta\Big)\, .
\end{equation}
This construction is valid provided that $N\ge 2$ and $M_i\ge 3$ for all $i$, which ensures that the total number of vertices $M$ satisfies the condition $M\ge 3N$. These requirements are indispensable for the construction of the proposed topological structure.

\section{Concluding Remarks} \label{sec:concluding}
The present work provides an analysis of the entanglement of quantum states associated with directed graph states, employing an entanglement measure derived from the Fubini–Study metric. We have shown that the vertex degree distribution solely determines the entanglement in this class of states and remains invariant under vertex relabeling. This invariance highlights the topological nature of the measure applied to quantum networks, emphasising that it is the topological properties of the graph —rather than the specific orientations of the edges— that govern the entanglement distribution. 


By establishing a direct link between graph topology and entanglement distribution, an extension of our work to weighted graphs with non-homogeneous vertex states could provide a geometric framework for the design and analysis of quantum networks \cite{verdon2019quantumgraphneuralnetworks,liao2024graphneuralnetworksquantum,ceschini2024graphsqubitscriticalreview}. This perspective may open new opportunities for integrating entanglement and network topology within quantum machine learning. In this setting, rather than relying on a fixed network structure, the learning process could be guided by the dynamic maximization of entanglement, allowing the topology to evolve in real time during training, with the aim of enhancing adaptability and improving overall efficiency.

A natural extension of this work is the study of decoherence effects by modelling the environment as a source of stochastic rearrangements of the graph connections. This approach could offer new insights into the stability of entanglement under topological perturbations, potentially contributing to the development of more resilient quantum communication protocols and fault-tolerant quantum operations. One possible strategy to mitigate decoherence involves exploiting the quantum Zeno effect, which could be used to control and stabilise entanglement near a desired value. The ability to manage and preserve entanglement is crucial not only for ensuring the robustness of quantum systems but also for applications in quantum metrology, where entanglement serves as a key resource for achieving high-precision measurements.

\appendix
\section{}
\label{appendix}
The ED per qubit \eqref{final} is derived in \cite{desimone2025entanglementdirectedgraphstates} by assuming that all vertices start from the commonly used state $\ket{\phi}=\frac{1}{\sqrt{2}}(\ket{0}+\ket{1})$, in line with the standard literature on graph states. As shown in \ref{sec:gs}, this state maximizes both the Entanglement Distance and the von Neumann entropy, while at the same time minimizing the Hilbert–Schmidt distance between $\ket{\phi}$ and the maximally mixed state $\mathbb{I}/2$. In this Appendix, we derive the general expression for the ED arising from the $i$-th vertex, by considering an arbitrary single-qubit input state $|\tilde{\phi}\rangle=\alpha_0\ket{0}+\alpha_1\ket{1}$, with $|\alpha_0|^2+|\alpha_1|^2=1$. Following the calculations of case \emph{iii)} in \cite{desimone2025entanglementdirectedgraphstates}, where the vertex $i$ is connected to $d_{\rightarrow}(i)$ vertices, $j_1,\dots,j_{d_{\rightarrow}(i)}\in\Gamma_{\rightarrow}(i)$, by $d_{\rightarrow}(i)$ outgoing links, and to $d_{\leftarrow}(i)$ vertices, $m_1,\dots,m_{d_{\leftarrow}(i)}\in\Gamma_{\leftarrow}(i)$, by $d_{\leftarrow}(i)$ incoming links.
Let us set $z=\langle\tilde{\phi}|\bar{U}|\tilde{\phi}\rangle$. The complete unitary operator results from the product of the operators \eqref{5}, with the corresponding adjustments in the indices, and is given by 
\begin{equation}
    U_{tot}=\prod^{d_{\rightarrow}(i)}_{k=1} {U}_{ij_k}\prod^{d_{\leftarrow}(i)}_{p=1}{U}_{m_p i}\, ,
\end{equation}
where, for the following calculations, we denote $U_{\rightarrow}$ and $U_{\leftarrow}$ as $\prod^{d_{\rightarrow}(i)}_{k=1} {U}_{ij_k}$ and $\prod^{d_{\leftarrow}(i)}_{p=1}{U}_{m_p i}$ respectively.

By direct calculation we get 
\begin{equation}
    \begin{split}
\langle\tilde{\phi}|^{d_{\rightarrow}(i)}&U_{\rightarrow}^{\dagger}\sigma^{(i)}_{\gamma}U_{\rightarrow}|\tilde{\phi}\rangle^{d_{\rightarrow}(i)} = \\
     &\ket{0}\bra{1}^{(i)}z^{d_{\rightarrow}(i)}+\ket{1}\bra{0}^{(i)}z^{*\,d_{\rightarrow}(i)}\\
     &-i\ket{0}\bra{1}^{(i)}z^{d_{\rightarrow}(i)}+i\ket{1}\bra{0}^{(i)}z^{*\,d_{\rightarrow}(i)}\\
    \end{split}
\end{equation}
for $\gamma=x,y$, and 
\begin{equation}
\langle\tilde{\phi}|^{d_{\rightarrow}(i)}U_{\rightarrow}^{\dagger}\sigma^{(i)}_{z}U_{\rightarrow}|\tilde{\phi}\rangle^{d_{\rightarrow}(i)} = 
\sigma_z^{(i)} \, .
\end{equation}
From the three relations above, one can directly calculate
\begin{equation}
\begin{split}
   &\langle\tilde{\phi}|^{\otimes M}
    {U}^{\dagger}_{tot}{\sigma}_x^{(i)}{U}_{tot}
    |\tilde{\phi}\rangle^{\otimes M} = \nonumber \\ \nonumber
    =&\langle\tilde{\phi}|^{\!(i)}\!
    \langle\tilde{\phi}|^{d_{\leftarrow}\!(i)}\! 
     U^\dagger_{\!\leftarrow}\!
     \langle\tilde{\phi}|^{d_{\rightarrow}\!(i)}\!
     U_{\!\rightarrow}^{\dagger}\!\sigma^{(i)}_{x}\!U_{\rightarrow}\!|\tilde{\phi}\rangle^{d_{\rightarrow}\!(i)}\!
    U_{\!\leftarrow}\!|\tilde{\phi}\rangle^{d_{\leftarrow}\!(i)}\!|\tilde{\phi}\rangle^{\!(i)}=\\
    =& 
    2 \Re{\sum_{k=0}^{d_{\leftarrow}(i)}\mathcal{B}(k;d_{\leftarrow}(i),p)\,\alpha_0^*\alpha_1 z^{d_{\rightarrow}(i)}e^{-i(d_{\rightarrow}(i)\psi+2k\theta)}}\, ,
\end{split}
\end{equation} 
\begin{equation}
\begin{split}
   &\langle\tilde{\phi}|^{\otimes M}
    {U}^{\dagger}_{tot}{\sigma}_y^{(i)}{U}_{tot}
    |\tilde{\phi}\rangle^{\otimes M} = \nonumber \\ \nonumber
    =&\langle\tilde{\phi}|^{\!(i)}\!
    \langle\tilde{\phi}|^{d_{\leftarrow}\!(i)}\! 
     U^\dagger_{\!\leftarrow}\!
     \langle\tilde{\phi}|^{d_{\rightarrow}\!(i)}\!
     U_{\!\rightarrow}^{\dagger}\!\sigma^{(i)}_{y}\!U_{\rightarrow}\!|\tilde{\phi}\rangle^{d_{\rightarrow}\!(i)}\!
    U_{\!\leftarrow}\!|\tilde{\phi}\rangle^{d_{\leftarrow}\!(i)}\!|\tilde{\phi}\rangle^{\!(i)}=\\
    =& 
    2\Im{\sum_{k=0}^{d_{\leftarrow}(i)}\mathcal{B}(k;d_{\leftarrow}(i),p)\,\alpha_0^*\alpha_1 z^{d_{\rightarrow}(i)}e^{-i(d_{\rightarrow}(i)\psi+2k\theta)}} \, ,
\end{split}
\end{equation} 
and
\begin{equation}
\begin{split}
   &\langle\tilde{\phi}|^{\otimes M}
    {U}^{\dagger}_{tot}{\sigma}_z^{(i)}{U}_{tot}
    |\tilde{\phi}\rangle^{\otimes M} = \nonumber \\ \nonumber
    =&\langle\tilde{\phi}|^{\!(i)}\!
    \langle\tilde{\phi}|^{d_{\leftarrow}\!(i)}\! 
     U^\dagger_{\!\leftarrow}\!
     \langle\tilde{\phi}|^{d_{\rightarrow}\!(i)}\!
     U_{\!\rightarrow}^{\dagger}\!\sigma^{(i)}_{z}\!U_{\rightarrow}\!|\tilde{\phi}\rangle^{d_{\rightarrow}\!(i)}\!
    U_{\!\leftarrow}\!|\tilde{\phi}\rangle^{d_{\leftarrow}\!(i)}\!|\tilde{\phi}\rangle^{\!(i)}=\\
    =& 
    |\alpha_0|^2-|\alpha_1|^2 \, .
\end{split}
\end{equation} 
Here, $\mathcal{B}(k;d_{\leftarrow}(i),p)$ denotes the binomial probability mass function $\mathcal{B}(k;d_{\leftarrow}(i),p)=\binom{d_{\leftarrow}(i)}{k}p^k (1-p)^{d_{\leftarrow}(i)-k}$. Expressing $z$, $\alpha_0$, and $\alpha_1$ as $z=r\,e^{i\delta}$, $\alpha_0=\sqrt{1-p} e^{i\delta_0}$ and $\alpha_1=\sqrt{p}e^{i\delta_1}$, we obtain the following expectation values of the Pauli operators
\begin{equation}
\langle\boldsymbol{\sigma}^{(i)}\rangle =
\begin{pmatrix}
2\sqrt{p(1-p)}\,r^{d(i)}\cos(\Phi) \\
-2\sqrt{p(1-p)}\,r^{d(i)}\sin(\Phi) \\
1-2p
\end{pmatrix}
\end{equation}
where $r=\smash{\sqrt{\,\cos^2(\theta)+\sin^2(\theta)(1-2p)^2}\,}$ and $\Phi=\delta_0-\delta_1-d(i)\delta+d_{\rightarrow}(i)\psi+d_{\leftarrow}(i)\theta$. The contribution of the $i$-th vertex to the Entanglement Distance is then given by
\begin{equation}
    E^{(i)}=1-(1-2p)^2-4p(1-p)r^{2d(i)}.
\end{equation}

This result shows that, even for a generic input state, the Entanglement Distance depends only on the degree distribution, thus preserving its purely topological character.

\begin{acknowledgments}
We acknowledge the support of the Research Support Plan 2022 – Call for applications for funding allocation to research projects curiosity-driven (F CUR) – Project "Entanglement Protection of Qubits’ Dynamics in a Cavity" – EPQDC and the support from the Italian National Group of Mathematical Physics (GNFM-INdAM). R. F. would like to acknowledge INFN Pisa for the financial support for this activity.
\end{acknowledgments}

\nocite{*}
\bibliography{references}

\end{document}